\documentclass[aps,amsfonts,amsmath,superscriptaddress,twocolumn,showpacs,floatfix,nofootinbib,showkeys]{revtex4-1}

\usepackage{marginnote}

\usepackage{slashed}
\usepackage{color, verbatim}
\usepackage{latexsym}
\usepackage{epsfig}
\usepackage{amsmath}

\usepackage{stackrel}

\usepackage{amssymb}
\usepackage{graphicx}
\usepackage{bm}

\usepackage{hyperref}
\usepackage{epstopdf}
\definecolor{myred}{rgb}{0.7, 0, 0}
\definecolor{myblue}{rgb}{0, 0, 0.7}
\definecolor{mygreen}{rgb}{0.04, 0.7, 0.5}
\hypersetup{colorlinks,citecolor=myred,linkcolor=myblue,urlcolor=myblue,linktocpage=true}

\makeatletter

\newcommand{\be}{\begin{equation}}
\newcommand{\ee}{\end{equation}}
\newcommand{\bea}{\begin{eqnarray}}
\newcommand{\eea}{\end{eqnarray}}

\newcommand{\GW}{\textrm{GW}}

\begin{document}

\title{Gravitational Imprints from Heavy Kaluza-Klein Resonances}

\author{Eugenio Meg\'{\i}as}
\email{emegias@ugr.es}
\affiliation{Departamento de F\'{\i}sica At\'omica, Molecular y Nuclear and  Instituto Carlos I de F\'{\i}sica Te\'orica y Computacional, Universidad de Granada, Avenida de Fuente Nueva s/n,  18071 Granada, Spain}

\author{Germano Nardini}
\email{germano.nardini@uis.no}
\affiliation{Faculty of Science and Technology, University of Stavanger, 4036 Stavanger, Norway}

\author{Mariano Quir\'os\,}
\email{quiros@ifae.es}
\affiliation{Institut de F\'{\i}sica d'Altes Energies (IFAE), The Barcelona Institute of  Science and Technology (BIST), Campus UAB, 08193 Bellaterra (Barcelona) Spain}

\date{\today}

\begin{abstract} 
\noindent We systematically study the holographic phase transition of the radion field in a five-dimensional warped model which includes a scalar potential with a power-like behavior. We consider Kaluza-Klein (KK) resonances with masses $m_{\rm KK}$ at the TeV scale or beyond. The backreaction of the radion field on the gravitational metric is taken into account by using the superpotential formalism. The confinement/deconfinement first order phase transition leads to a gravitational wave stochastic background which mainly depends on the scale $m_{\rm KK}$ and the number of colors, $N$, in the dual theory. Its power spectrum peaks at a frequency that depends on the amount of tuning required in the electroweak sector. It turns out that the present and forthcoming gravitational wave observatories can probe scenarios where the KK resonances are very heavy. Current aLIGO data already rule out vector boson KK resonances with masses in the interval $m_{\rm KK}\sim(1 - 10) \times 10^5$~TeV. Future gravitational experiments will be sensitive to resonances with masses $m_{\rm KK}\lesssim 10^5$~TeV~(LISA), $10^8$~TeV~(aLIGO Design) and $10^9$~TeV~(ET). Finally, we also find that the Big Bang Nucleosynthesis bound in the frequency spectrum turns into a lower bound for the nucleation temperature as $T_n  \gtrsim 10^{-4}\sqrt{N} \,m_{\rm KK}$. 
\end{abstract}



\maketitle



\section{Introduction}
\label{sec:Intro}

The Standard Model is unable to explain some experimental observations (e.g.~dark matter, the baryon asymmetry of the universe, \dots), and suffers from theoretical drawbacks (e.g.~strong sensitivity to high scale physics, a.k.a.~hierarchy problem, \dots). A warped extra dimension is a way of solving the hierarchy problem and relating the Planck scale $M_P$ to the low energy scale $\rho$, which determines the spectrum of heavy resonances and is usually considered at the TeV scale~\cite{Randall:1999ee,Goldberger:1999uk}.
 However, the elusiveness of experimental data on the search of stable narrow resonances~\cite{Sirunyan:2018ryr,Aaboud:2019roo} is perhaps suggesting us that nature might not be as generous as we assumed it to be, and is not solving the ``whole" hierarchy problem but only part of it, in which case $\rho$ can be much heavier than the TeV scale, worsening the little hierarchy problem. 

But, had nature chosen that way, where could we find sensitivity to such heavy physics, aside from future more energetic colliders? The answer is based on the presence of the only extra light field in the theory, the radion. This field experiences a first order phase transition, the confinement/deconfinement transition, which generates a stochastic gravitational wave background (SGWB) detectable at the present  and future interferometers~\cite{Caprini:2015zlo, Bartolo:2016ami, Caprini:2018mtu}. 
In this paper we cover this issue for the minimal five-dimensional (5D) warp model~\cite{Randall:1999ee} with a stabilizing field  with a bulk polynomial potential. Studies of the holographic phase transition have been performed with great detail in the literature~\cite{Creminelli:2001th,Randall:2006py,Nardini:2007me,Konstandin:2010cd,Konstandin:2011dr,Bunk:2017fic,Dillon:2017ctw,vonHarling:2017yew,Bruggisser:2018mus,Bruggisser:2018mrt,Megias:2018sxv,Baratella:2018pxi,Agashe:2019lhy,Fujikura:2019oyi}.
Here we make a step forward in several aspects: \textit{i)} We take into account the full backreaction of the scalar field on the gravitational metric, using the superpotential mechanism and methods proposed in Ref.~\cite{Megias:2018sxv}; \textit{ii)} We moreover go beyond the common beliefs on what is allowed by the little hierarchy problem, and thus explore parameter regions with large $\rho$, while still solving the big hierarchy between $M_P$ and $\rho$ by means of the metric warped factor.

The outline of the paper is as follows. Section~\ref{sec:model} introduces the considered warped model, some conventions, and the technique adopted to accurately treat the backreaction on the metric. Section~\ref{sec:effective_potential} deals with the radion effective potential and shows how the backreaction and detuning of the brane tensions impact it. Section~\ref{sec:radion_field} includes some key elements of the radion phenomenology. Section~\ref{sec:phase_transition} and Section~\ref{sec:GW}, respectively, deal with the radion phase transition and its gravitational wave signatures. Finally, Section~\ref{sec:conclusions} summarizes the main results and some remarks.

\section{The model}
\label{sec:model}

We consider a scalar-gravity system, with metric $g_{MN}$ defined in proper coordinates by 
\be
ds^2 = g_{MN}dx^M dx^N\equiv e^{-2A(r)} \eta_{\mu\nu} dx^\mu dx^\nu-dr^2\,,
\ee
 and two branes at $r = r_a$, where $a=0,1$ for the ultraviolet (UV) and infrared (IR) brane, respectively. We fix $r_0 = 0$ by convention, and our notation follows that in Ref.~\cite{Megias:2018sxv}. 

The five-dimensional action of the model reads as
\begin{align}
S &= \int d^5x \sqrt{|\det g_{MN}|} \bigg[ -\frac{1}{2\kappa^2} R + \frac{1}{2} g^{MN}(\partial_M \phi)(\partial_N \phi) \nonumber \\
-&V(\phi) \bigg] - \sum_{a} \int_{B_a} d^4x \sqrt{|\det \bar g_{\mu\nu}|} \Lambda_a(\phi) + S_{\rm GHY} \,, \label{eq:action}
\end{align}
where we have introduced a bulk scalar field with mass dimension $3/2$. There are three kind of contributions to the action, corresponding to the bulk, the brane, and the Gibbons-Hawking-York term. $V(\phi)$ (with mass dimension 5) and $\Lambda_\alpha(\phi)$ (with mass dimension 4) are the bulk and brane potentials of the scalar field~$\phi$, 
while the four-dimensional induced metric is $\bar g_{\mu\nu} = e^{-2 A(r)} \eta_{\mu\nu}$. For concreteness we will consider the brane potentials $\Lambda_a(\phi)$ as
\be
\Lambda_a(\phi)=\Lambda_a+\frac{1}{2}\gamma_a(\phi-v_a)^2 \,,
\ee
where $\Lambda_a$ is a constant, hereafter considered as a free parameter, and $\gamma_a$ is a dimensionful parameter. We will also work in the stiff potentials limit, where $\gamma_a\to\infty$, such that the values of the bulk field at the branes are $\phi(r_a)=v_a$~\cite{Goldberger:1999uk}.

The background equations of motion (EoM) can be expressed in terms of the superpotential $W(\phi)$ (with mass dimension 4), as~\cite{DeWolfe:1999cp} 
\begin{align}
\phi^\prime(r)& = \frac{1}{2} W^\prime( \phi) \,, \qquad A^\prime(r) = \frac{\kappa^2}{6} W(\phi) \,, \nonumber\\
V(\phi) &= \frac{1}{8} \left[W^\prime( \phi) \right]^2 - \frac{\kappa^2}{6} W^2(\phi) \,, \label{eq:V}
\end{align}
where the prime symbol $(\,{}^\prime\,)$ stands for the derivative of a function with respect to its argument, and  $\kappa^2\equiv 1/(2 M^3)$,  $M$ being the 5D Planck mass.
$W$ is expressed as the expansion $W=\sum_n s^nW_n$~\cite{Papadimitriou:2007sj,Megias:2014iwa,Megias:2018sxv,Lizana:2019ath} where the parameter $s$ is dimensionless, and we are choosing
\be
W_0(\phi)=\frac{6}{\ell\kappa^2}+\frac{u}{\ell}\phi^2 \,,
\label{eq:W0}
\ee
with $\ell$ being an $\mathcal{O}(M^{-1})$ parameter, $u$ being dimensionless, and $s$ playing the role of the (small) integration constant of Eq.~(\ref{eq:V}). We work to linear approximation in~$s$ and keep the leading terms of the $u\ll 1$ limit, for which one can solve the hierarchy problem with $\mathcal O(1)$ values for $v_a$ in units of $\kappa^{-1}$.

To linear order, we have $W=W_0+sW_1$ with the $W_1$ component of the superpotential given by~\cite{Megias:2018sxv}
\be
W_1(\phi)=\frac{1}{\ell\kappa^2}\left(\frac{\phi}{v_0}\right)^{4/u}e^{\kappa^2(\phi^2-v_0^2)/3}  \,.
\label{eq:W1}
\ee
Similarly $\phi$ can be decomposed as $\phi=\phi_0+s \phi_1$ with~
\be
\bar\phi_0(r)=\bar v_0e^{u \bar r}
\ee
 and
\begin{eqnarray}
&& \bar\phi_1(r)=\frac{1}{2 u \bar v_0}e^{u \bar r}\left[ e^{(4-2 u)\bar r}e^{\bar v_0^2/3\left( e^{2 u\bar r}-1 \right)}-1  \right]\,,
\label{eq:phi1app}
\end{eqnarray}
which fulfills the UV boundary condition $\phi(r_0)=v_0$~\cite{Megias:2018sxv}. The IR boundary condition $\phi(r_1)=v_1$ instead requires
\begin{eqnarray}
&&s(\bar r_1)=\frac{2 u \bar v_0^2 e^{-u \bar r_1}\left( e^{u\bar r_1^0}-e^{u\bar r_1} \right)}{e^{(4-2 u)\bar r_1}e^{\bar v_0^2/3\left(e^{2 u \bar r_1}-1 \right)}-1}\,.
\label{eq:s}
\end{eqnarray}
Likewise the expansion of the metric exponent, $A(r)=A_0(r)+s A_1(r)$, yields
\begin{align}
A_0( r)&=\bar r+\frac{\bar v_0^2}{12}\left(e^{2 u  \bar r}-1  \right)  \,,
\label{eq:A0}\\
A_1(r)& = \frac{1}{12}\left[ e^{4A_0(\bar r)}-1\right] +\frac{2+u}{24 u}\left(1-\frac{\bar\phi_0^2}{\bar v_0^2} \right) \,.
\label{eq:A1}
\end{align}
For convenience, in the above expressions we have introduced the dimensionless quantities $
\bar v_a\equiv\kappa v_a,\ \bar\phi(r)\equiv\kappa\phi(r),\ \bar r\equiv r/\ell,\ \bar r_a\equiv r_a/\ell
$. 

\section{The effective potential}
\label{sec:effective_potential}

The effective potential normalized to its value at $r_1\to\infty$ is given, in the stiff limit for boundary potentials, by~\cite{Megias:2018sxv}
\begin{align}\label{eq:Ueff}
U_{\rm eff}(r_1) &=[\Lambda_1+W_0(v_1)]e^{-4 A_0(r_1)}[1-4A_1(r_1)s(r_1)] \nonumber \\
&+ s(r_1)\left[e^{-4 A_0(r_1)}W_1(v_1)-W_1(v_0)  \right] \,,
\end{align}
where $\Lambda_1$ is the tension at the IR brane. From the expression of the effective potential given in Eq.~(\ref{eq:Ueff}) we can see that, even if the superpotential is very appropriate a tool to take into account the backreaction on the metric, to zeroth order in the expansion parameter $s$ it provides no fixing of the brane distance, as its dependence on $r_1$ through $A_0(r_1)$ yields a runaway behavior. Therefore, in order for the effective potential to fix the brane distance~$r_1$, we need to go to, at least, first order in the expansion of the parameter $s(r_1)$. In what follows, the smallness of $s(r_1)$ in the considered region of parameters indeed justifies truncating the series expansion to first order.

\subsection{The tuned potential}
We can tune to zero the first term of Eq.~(\ref{eq:Ueff}) by fixing  
\be
\Lambda_1=-W_0(v_1)\,.
\label{eq:tuned}
\ee
 Consequently the leading-order dimensionless effective potential 
 \be
 \bar U_{\rm eff}^0(r_1)\equiv \ell\kappa^2 U_{\rm eff}^0(r_1)
 \ee
  is given by 
\begin{align}
&\bar U_{\rm eff}^0(\bar r_1)=2 u \bar v_0^2\, e^{-u \bar r_1} \left[ e^{(4-2 u) \bar r_1}e^{\bar v_0^2/3\left( e^{2 u \bar r_1}-1 \right)}-1  \right]^{-1}  \nonumber \\
&   \left[e^{u \bar r_1^0}-e^{u \bar r_1} \right]\left[e^{-4(\bar r_1-\bar r_1^0)}e^{\bar v_0^2/3\left( e^{2 u \bar r_1^0}-e^{2 u \bar r_1} \right)}-1 \right] \;,
\label{eq:Uefftuned}
\end{align}
where $\bar r_1^0$ is defined by the condition
\be
v_1\equiv v_0e^{u \bar r_1^0} \,.
\label{eq:r10}
\ee
In fact, an excellent approximation for the tuned effective potential is given by
\be
\bar U_{\rm eff}^0(\bar r_1)=2 u^2 \bar v_1^2 (\bar r_1^0-\bar r_1)
\left[e^{4A_0(\bar r_1^0)-4A_0(\bar r_1)}-1\right]\, e^{-4 A_0(\bar r_1)}  . \label{eq:Uefftuned_approx}
\ee
Notice that the expression of the effective potential in Eqs.~(\ref{eq:Uefftuned}) and (\ref{eq:Uefftuned_approx}) vanishes when $u=0$, i.e.~in the absence of backreaction. Note also that $A_0(\bar r_1)$ is a positive increasing function for $\bar r_1 > 0$, and thus the factor $(\bar r_1^0 - \bar r_1)$ and the term inside the bracket have the  same sign for any~$r_1$. The potential in Eq.~(\ref{eq:Uefftuned_approx}) is therefore positive definite. Moreover, one can see that $\bar U_{\rm eff}^0(\bar r_1)$ has degenerate minima, at $\bar r_1=\bar r_1^0$ and $\bar r_1\to\infty$, where it vanishes. 
\begin{figure*}[htp]
\centering
 \begin{tabular}{c@{\hspace{2.5em}}c}
 \includegraphics[width=0.43\textwidth]{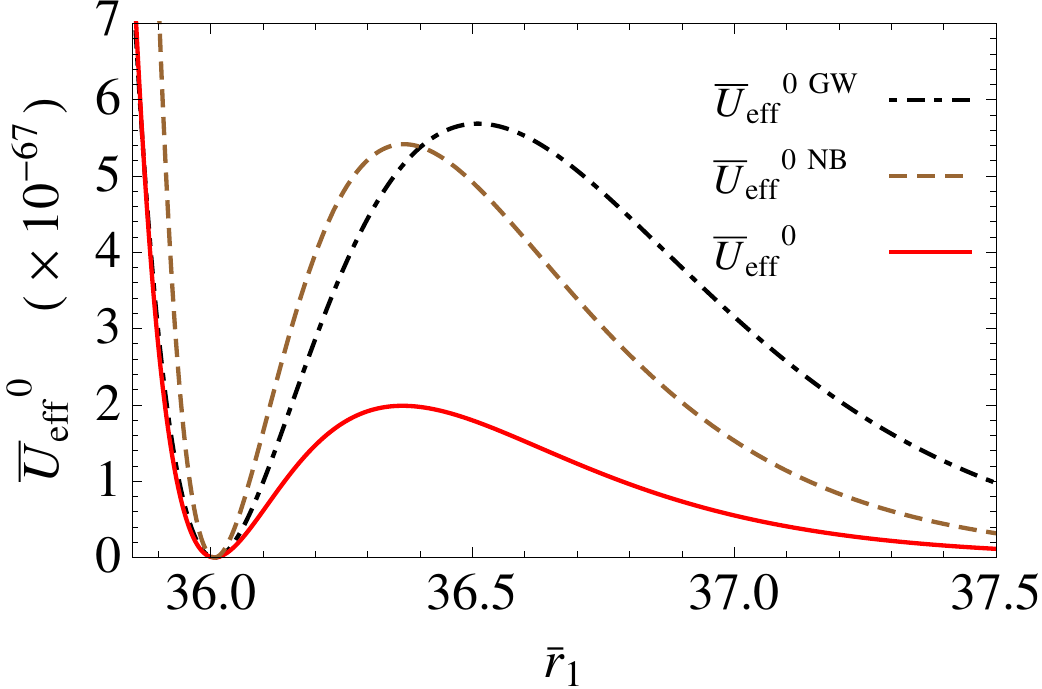} &
 \includegraphics[width=0.46\textwidth]{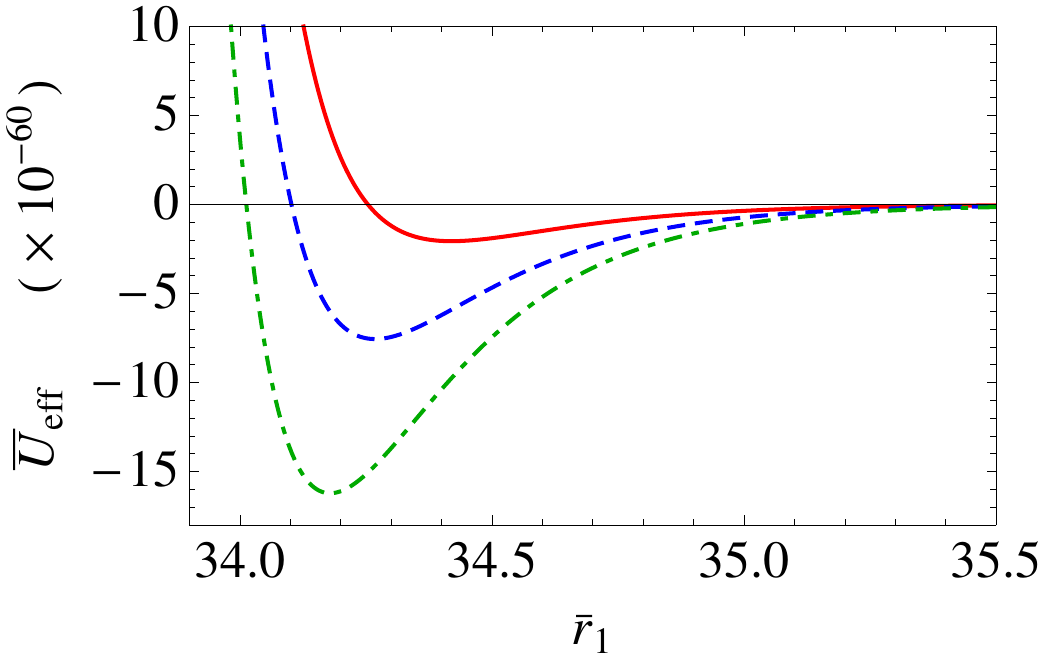} 
\end{tabular}
\caption{\it Left panel: The potential $\bar U_{\rm eff}^0$ in
  Eq.~(\ref{eq:Uefftuned_approx}) (solid line), the same potential
  when neglecting the backreaction $\bar U_{\rm eff}^{0 \rm NB}$
  (dashed line), and the Goldberger-Wise potential $\bar U_{\rm eff}^{0 \rm GW}$ in
  Eq.~(\ref{eq:UGW}) (dashed-dotted line) as functions of~$\bar
  r_1$. Right panel: The potential $\bar U_{\rm eff}$ in
  Eq.~(\ref{eq:Ueffdetuned}) for
  $\lambda_1=$-1 (red solid line), -2 (blue dashed line) and -3 (green
  dashed-dotted line). In both panels we assume  $\bar
  v_0=1,\, \bar v_1=2,\, u=0.0192$\,.}
\label{fig:Ueff}
\end{figure*}
For the sake of comparison, the Goldberger-Wise potential of Ref.~\cite{Goldberger:1999uk} can be written in our notation as
\be
\bar U_{\rm eff}^{0 \, \rm GW}(\bar r_1) = 4 \bar v_0^2 \, e^{-4 \bar r_1} \left( e^{u \bar r_1^0 } - e^{-u \bar r_1} \right)^2 +\mathcal O(u)\; ,
\label{eq:UGW}
\ee
which is positive definite when ignoring the $\mathcal{O}(u)$ terms.

The left panel of Fig.~\ref{fig:Ueff} displays the plot of the
effective potential in Eq.~(\ref{eq:Uefftuned_approx}) for parameter
values leading to a potential minimum at $\bar r_1^0=36$. The
expansion hierarchy $s(\bar r_1)W_1(\bar r_1)/W_0(\bar
r_1)\sim\mathcal O(10^{-4})$, so the $s$-expansion of the
superpotential converges fast. For the sake of comparison, the left
panel of Fig.~\ref{fig:Ueff} also shows the Goldberger-Wise potential
$\bar U_{\rm eff}^{0 \, \rm GW}$, as well as the potential
(\ref{eq:Uefftuned_approx}) with an artificially neglected
backreaction on the metric, $\bar U_{\rm eff}^{0 \,\rm NB}$,
i.e.~considering $A_0(r) \simeq \bar r$. We can see in the left panel
of Fig.~\ref{fig:Ueff} that both potentials, $\bar U_{\rm eff}^{0 \,
  \rm GW}$ and $\bar U_{\rm eff}^{0 \, \rm NB}$, agree very well, even
if they have been computed using completely different methods.

\subsection{The detuned potential}
Due to the above ``tuning'', the potential minima at $\bar r_1=\bar r_1^0$ and $\bar r_1=+\infty$ are degenerate. This prevents the required transition from the deconfined to the confined phase in the early universe. In order to allow such a phase transition, we need to detune condition (\ref{eq:tuned}), which leads to the potential in Eq.~(\ref{eq:Uefftuned}), and introduce a non-vanishing parameter $\lambda_1$ as 
\be
\Lambda_1+W_0(v_1)=\Lambda_1+\frac{6+u \bar v_1^2}{\ell\kappa^2}\equiv\frac{6}{\ell\kappa^2}\lambda_1\neq 0 \,,
\label{eq:detuned}
\ee
with $\lambda_1$ a dimensionless parameter. In this case the dimensionless effective potential is given by
\begin{align}
\bar U_{\rm eff}(\bar r_1) &\simeq\bar U^0_{\rm eff}(\bar r_1)+6\lambda_1 e^{-4 A_0(\bar r_1)} \,,
\label{eq:Ueffdetuned}
\end{align}
where the term $4 A_1(\bar r_1) s(\bar r_1)$ has been
omitted since it is smaller than $10^{-3}$ in the parameter region we focus on. In the right panel of Fig.~\ref{fig:Ueff} we plot the
effective potential in Eq.~(\ref{eq:Ueffdetuned}) for various values
of $\lambda_1$. For $\lambda_1<0$ the global minimum is at a finite
value of $r_1$. For positive values of $\lambda_1$ the minimum at
finite values of $r_1$ is not the global minimum, or just
disappears. 

The position of the minimum of $\bar U_{\rm eff}^0(r_1)$
in Eq.~(\ref{eq:Uefftuned_approx}), $\bar r_1^0$, and the true minimum of $\bar U_{\rm eff}(\bar r_1)$
in Eq.~(\ref{eq:Ueffdetuned}) differ by a small amount $\delta$. By
denoting the latter as $\bar r_1^{\textrm{m}}$ and plugging the
difference 
\be
\delta = \bar r_1^\textrm{m} - \bar r_1^0
\ee
into Eq.~(\ref{eq:Ueffdetuned}), one finds
\begin{eqnarray}
&&\delta\simeq -\frac{1}{4}\mathcal W\left[-\frac{6\lambda_1}{u^2 \bar v_1^2}\right]  \,.
\label{eq:Lambert}
\end{eqnarray}
Here $\mathcal W$ is the Lambert $\mathcal W$ function~\footnote{The Lambert function $\mathcal W(z)$ is defined as the principal solution (upper branch) for the equation $\mathcal W e^{\mathcal W}=z$.}. The approximation leading to Eq.~(\ref{eq:Lambert}) relies on the expansion $u \ll 1$ and holds within a few per mille.

\section{The radion field}
\label{sec:radion_field}
Using the formalism of Ref.~\cite{Megias:2018sxv} we find, to leading approximation in the parameter~$u$, that the radion field 
$\bar\chi(r_1)\equiv \ell\chi(r_1)$ can be approximated by
\be
\bar\chi(r_1)\simeq e^{-A_0(r_1)} \,,
\ee
an expression which can be inverted, and yields
\be
\bar r_1(\bar\chi)=-\log\bar\chi+\frac{\bar v_0^2}{12}-\frac{1}{2u}\mathcal W\left[\frac{u \bar v_0^2}{6}e^{u (\bar v_0^2/6 -2\log\bar\chi)} \right] \,.
\ee

\begin{figure}[tb]
\centering 
\includegraphics[width=7cm]{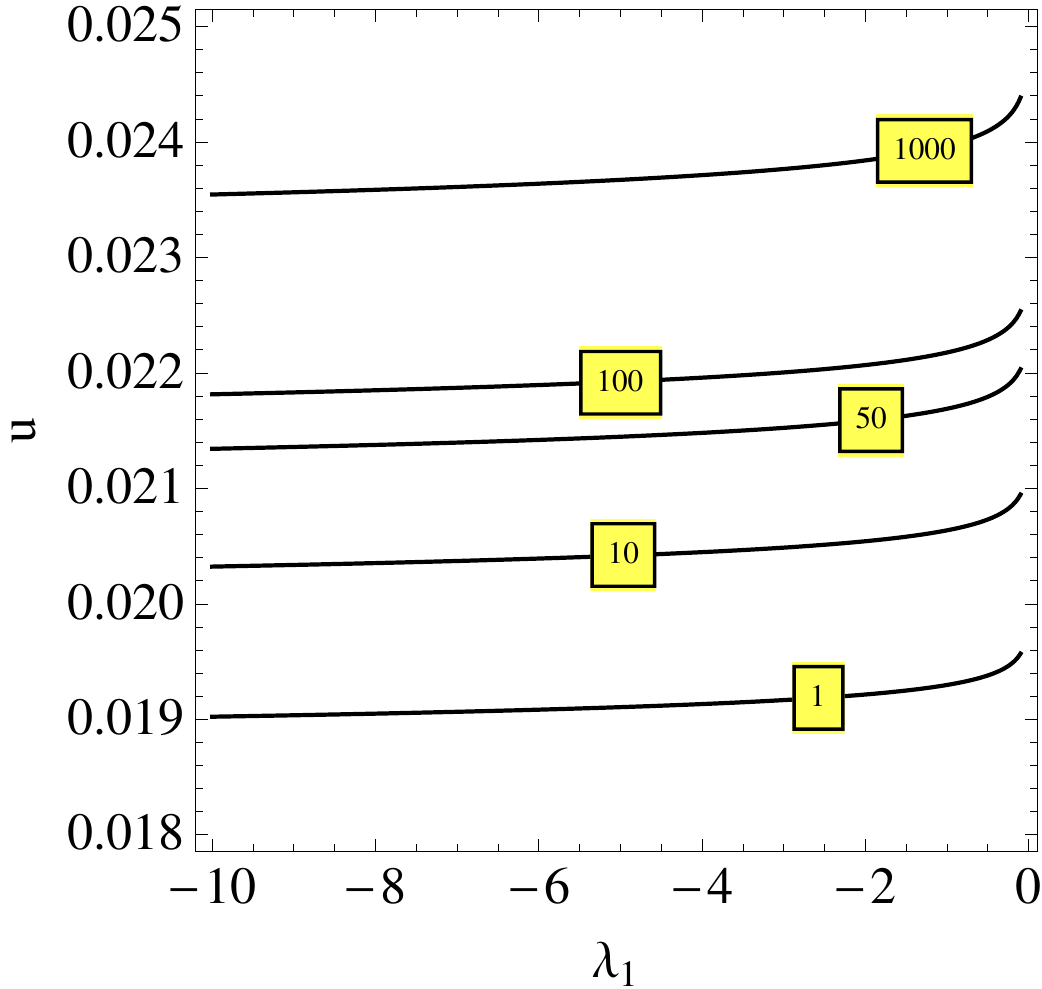} 
\caption{\it Contour plot for values of $\rho$ in TeV units, for $\bar v_0 = 1$, $\bar v_1 = 2$ and $1/\ell = 10^{18}$~GeV.}
\label{fig:rho}
\end{figure}

We now introduce the physically relevant parameter $\rho$ as
\be
\rho \equiv e^{-A_0(\bar r_1^m)}/\ell\,.
\ee
Contour lines of $\rho$ (in TeV) are exhibited in Fig.~\ref{fig:rho} in the plane $(\lambda_1,u)$ for the specified values of the parameters. We see from Fig.~\ref{fig:rho} that
 $\rho$ is
mainly determined by $u$, with a milder dependence
on~$\lambda_1$. We find $u\simeq 0.0192 \,
(0.0219)$ for $\rho=1\, (100)$~TeV as a set of benchmark values, although we will use the precise functional dependence of $u$ on $\lambda_1$,  provided by Fig.~\ref{fig:rho}, in the rest of our numerical analysis.

A convenient parametrization of the (dimensionful) effective potential in units of the physically relevant parameter $\rho$ is then
\be
U_{\rm eff}(\bar r_1)=\frac{N^2\rho^4}{8\pi^2} e^{4 A_0(\bar r_1^m)}\bar U_{\rm eff}(\bar r_1) \,,
\label{eq:potredefinido}
\ee
where we are using the precise AdS/CFT relation on the 5D squared gravitational coupling constant $(M\ell)^{-3}$
\be
\frac{1}{N^2}=\frac{(M\ell)^{-3}}{16\pi^2},\quad \textrm{i.e.}\quad
N^2=8\pi^2 \ell^3 /\kappa^2\,,
\label{eq:AdSCFT}
\ee
 $N$ being the number of colors in the dual theory, as a ``definition'' of $N$. Hence the radion potential, given by 
\be
V_{\rm rad}(\chi)\equiv U_{\rm eff}[\bar r_1(\bar\chi)]\,,
\ee has a minimum at $\langle\chi\rangle=\rho$.

We now compute the radion mass using the mass formula of Ref.~\cite{Megias:2018sxv}. In the stiff limit for brane potentials, for which the radion mass is maximized, we can write
\be
m_{\rm rad}^2=  \rho^2/\Pi_{\rm rad}
\ee
 and 
\begin{align}
&\Pi_{\rm rad}=\frac{1}{\ell^2}\int_0^{r_1^m}dr e^{4(r-r_1^m)/\ell}e^{4(\Delta A(r)-\Delta A(r_1^m)}\left(\frac{W[\phi(r)]}{W'[\phi(r)]} \right)^2\nonumber\\
&\times \left[   \frac{2}{W[\phi(r_1^m)]}+\int_r^{r_1^m}d\bar r e^{-2(A(\bar r)-A(r_1^m))} \left(\frac{W'[\phi(\bar r)]}{W[\phi(\bar r)]} \right)^2
\right]
\label{eq:Pirad0}
\end{align}
with $\Delta A(r)\equiv A(r)-\bar r$. The observation that the integral in Eq.~(\ref{eq:Pirad0}) is dominated by the region $r\simeq r_1^m$ allows for an analytical approximation of the integral. Under such an approximation, in the limit $u\ll 1$ we obtain
\begin{align}
m_{\rm rad}/\rho&\simeq \left. e^{2[\Delta A(\bar r_1^m)- \Delta A(r)]} f(\bar r)\right|_{\bar r=\bar r_1^m-1/4} \,, \nonumber\\
f(\bar r)&=\sqrt{2\ell\,W[\phi(r_1^m)]}\frac{W'[\phi(r)]}{W[\phi(r)]} \,.
\label{eq:masaradionapp}
\end{align}

\begin{figure}[t]
\centering 
\includegraphics[width=7cm]{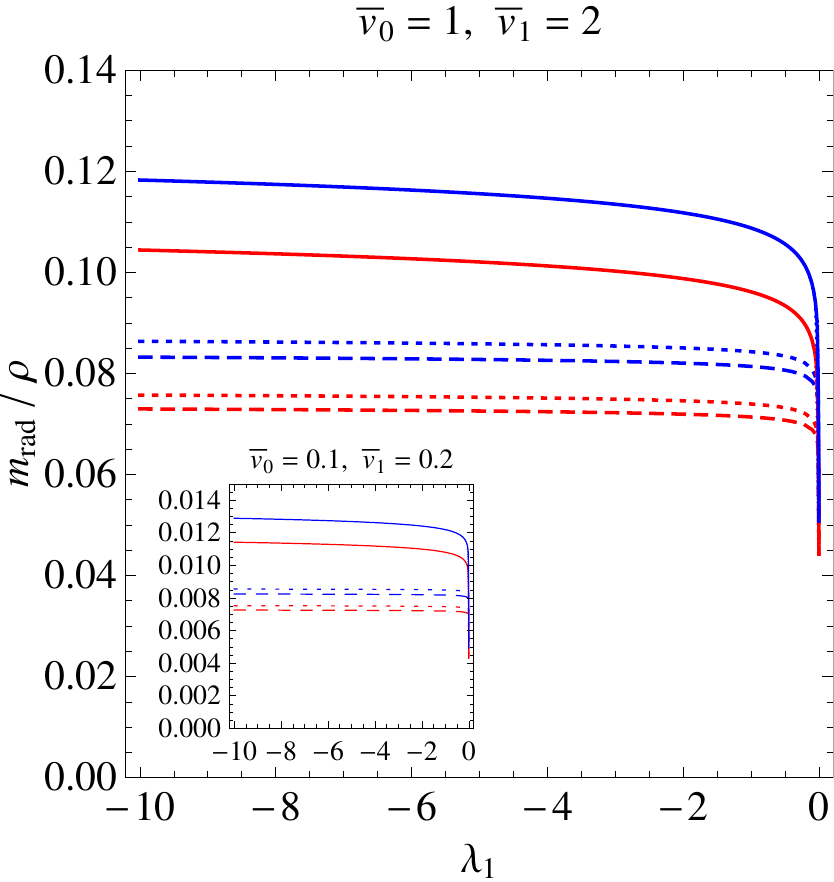} 
\caption{\it The normalized radion mass as a function of $\lambda_1$ for $\bar v_0 = 1,\, \bar v_1 = 2$ (main plot), and for $\bar v_0 = 0.1,\, \bar v_1 = 0.2$ (inserted figure). The red (blue) lines correspond to the cases $\rho = 1$ TeV (100 TeV). Solid lines are the radion mass from an exact numerical solution of the EoM.  The results of the mass formula in Eq.~(\ref{eq:Pirad0}), evaluated numerically are the dashes lines, and that from the approximation in Eq.~(\ref{eq:masaradionapp}), the dotted lines.}
\label{fig:radionmass}
\end{figure} 

We display in Fig.~\ref{fig:radionmass} the normalized radion mass as
a function of $\lambda_1$ by using: i) Dashed lines: the numerical
computation of the mass formula of Eq.~(\ref{eq:Pirad0}); ii) Dotted
lines: the analytical formula of Eq.~(\ref{eq:masaradionapp}); and,
iii) Solid lines: the numerical solution of the EoM of the scalar
perturbations along the lines of Ref.~\cite{Megias:2015ory}. Note that
the three methods are in reasonable agreement. The result for $\bar
v_0=1$, $\bar v_1=2$ and $\rho=1\,(100)$\, TeV is $m_{\rm
  rad}/\rho\simeq 0.10$ (0.12). For $\bar v_0=0.1,\, \bar v_1=0.2$,
where the backreaction on the metric is smaller, the radion is much
lighter, $m_{\rm rad}/\rho\simeq 0.011$ (0.013) with
$\rho=1\,(100)$\,TeV, as expected. In summary, we find that the radion
mass scales linearly with the values of $\bar v_0$ and $\bar v_1$,
while it is almost independent of $\lambda_1$.

Using similar approaches, we 
compute the mass of the first KK resonance for gauge bosons, $m_{\rm KK}^{\rm gauge}$, and
gravitons, $m_{\rm KK}^{\rm grav}$. We obtain $m_{\rm KK}^{\rm gauge}/\rho \simeq 2.46\,
(2.43)$ and $m_{\rm KK}^{\rm grav}/\rho \simeq 3.88\, (3.83)$ for
$\bar v_0=1,\, \bar v_1=2$ $(\bar v_0 = 0.1,\, \bar v_1 = 0.2)$, values almost independent of $\lambda_1$ and $\rho$.

For concreteness, hereafter we will consider the case $\bar v_0=1\,$, $\bar v_1=2$ and
$\ell$ nearby the Planck length $\ell_P$, namely
$1/\ell=10^{18}$\,GeV$\simeq 0.4\,/\ell_P$.

\section{The confinement/deconfinement phase transition}
\label{sec:phase_transition}
It is the phase transition from the radion symmetric (deconfined) phase, at $\chi=0$, to its broken (confined) phase, at $\chi=\langle\chi\rangle\neq 0$. At finite temperature the warped model admits an additional gravitational solution with a black hole (BH) singularity located at the event horizon $r=r_h$~\cite{Witten:1998zw,Creminelli:2001th},
\begin{equation}
ds^2_{\rm BH}= - h(r)^{-1} dr^2+e^{-2A(r)}(h(r)dt^2-d\vec x^2)\,.
\end{equation}
Here $h(r)$ is the blackening factor satisfying the boundary and regularity conditions $h(0)=1$ and $h(r_h)=0$.

A solution of the EoM~\cite{Megias:2018sxv} provides the function $h(r)$, in the $u\ll 1$ limit, as
\be
h(r)\simeq 1-e^{4[A_0(r)-A_0(r_h)]}\,,
\ee
which translates into $\ell h^\prime(r_h)\simeq -4$ since $A_0^\prime(r)=1+\mathcal  O(u)$. Thus the Hawking temperature, $T_h$, and the minimum of the free energy in the BH solution, at $T_h=T$, 
$F^{\rm BH}(T_h)$, read as
\be
\ell T_h\equiv \bar T_h\simeq \frac{1}{\pi}e^{-A_0(r_h)} \,,
\quad
F_{\rm min}^{\rm BH}(T)\simeq -\frac{\pi^4\ell^3}{\kappa^2}T^4\,.
\label{eq:Fmin}
\ee
In fact the free energy in the deconfined, $F_d(T)$, and confined, $F_c(T)$, phases at high temperature are given by
\begin{align}
F_d(T)&=E_0+F_{\rm min}^{\rm BH}(T)-\frac{\pi^2}{90}g_d^{\rm eff}T^4 \,, \nonumber\\
F_c(T)&=-\frac{\pi^2}{90}g_c^{\rm eff}T^4 \,,
\label{eq:free}
\end{align}
where $g_{d/c}^{\rm eff}$ is the number of relativistic degrees of freedom in the deconfined/confined phase, and $E_0=V_{\rm rad}(0)-V_{\rm rad}(\rho)$ is the potential gap between the two phases in the $T=0$ limit.

Below the critical temperature $T_c$, defined by 
\be
F_c(T_c)=F_d(T_c)\,,
\ee
 the phase transition can start.
We will assume $g_{d}^{\rm eff}(T_c) \simeq g_{c}^{\rm eff}(T_c)$~\footnote{This approximation holds e.g.~in setups with the right-handed top and Higgs localized on the IR brane and the remaining Standard Model fermions being elementary. Indeed in the energy budget the difference between $g_d^{\rm eff}(T_c)=97.5$ and $g_c^{\rm eff}(T_c)=106.75$ is negligible as compared to $45 N^2/4$.}, and
thus the critical temperature can be estimated as~
\be
\pi T_c/\rho\simeq e^{A_0(\bar r_1^m)} |\bar U_{\rm eff}(\bar r_1^m)|^{1/4}\,.
\label{eq:tempcritica}
\ee
 We find that $T_c/\rho$ is mostly insensitive to the particular value of $u$ in the parameter region considered in this work.

The (first order) phase transition proceeds through bubble nucleation of the confined phase in the deconfined sea. The onset of the transition occurs at the nucleation temperature $T_n$, where $T_n < T_c$. To compute $T_n$ we compute the Euclidean actions and compare them with the expansion rate of the universe at the corresponding temperature. 

In particular, at high temperature, the Euclidean action $S_3/T$, with symmetry $O(3)$, is given by
\be
S_3=4\pi\int d\sigma \sigma^2\, \frac{3N^2}{4\pi^2}\left[ \frac{1}{2} \left(\frac{\partial\chi}{\partial \sigma}\right)^2+V(\chi,T)  \right],
\ee
where $\sigma\equiv\sqrt{\vec x^2}$ is the space radial coordinate, with potential
\be 
V(\chi,T)=\frac{4\pi^2}{3N^2} \left( V_{\rm rad}(\chi) + \left|F_{\rm min}^{BH}(T)\right|  \right) \,,
\ee
bounce equation for $\chi=\chi(\sigma)$
\be
\frac{\partial^2\chi}{\partial\sigma^2}+\frac{2}{\sigma}\frac{\partial\chi}{\partial\sigma}=\frac{\partial V}{\partial\sigma} \,,
\ee
initial condition $\chi_0\equiv \chi(0)$, as well as boundary conditions
\be
\frac{3N^2}{8\pi^2}\left( \frac{\partial\chi}{\partial\sigma} \right)^2_{\chi=0}=\left|F_{\rm min}^{BH}(T)\right|,\
\left.\frac{d\chi}{d\sigma}\right|_{\sigma=0}=0 \,.
\label{eq:BC}
\ee

At low temperature there is also the $O(4)$ symmetric solution with action $S_4$ given by
\be
S_4=2\pi^2\int d\sigma \sigma^3\,\frac{3N^2}{4\pi^2}\left[ \frac{1}{2} \left(\frac{\partial\chi}{\partial \sigma}\right)^2+V(\chi,T)  \right],
\ee
where $\sigma=\sqrt{\vec x^2+\tau^2}$ ($\tau$ is the Euclidean time), and with bounce equation
\be
\frac{\partial^2\chi}{\partial\sigma^2}+\frac{3}{\sigma}\frac{\partial\chi}{\partial\sigma}=\frac{\partial V}{\partial\sigma} \,,
\ee
and boundary conditions given in Eq.~(\ref{eq:BC}).

Although we expect, on general grounds, the bubble formation to be dominated by thick wall approximation, at least for $T_n\ll T_c$~\cite{Randall:2006py,Bunk:2017fic}, we find that this approximation often mismatches the fully numerical result. We thus compute $S_3/T$ and $S_4$ via the numerical methods introduced in~\cite{Konstandin:2010cd,Megias:2018sxv}, and subsequently obtain $T_n$ from the condition
\be
S_E(T_n)\simeq 4 \log \frac{M_p/\rho}{T_n/\rho} 
\label{eq:criterium}
\ee
with
\be
S_E\equiv\min\left[\frac{S_{3}(T_n)}{T_n},S_4(T_n)\right]\,. \label{eq:SE} 
\ee

\begin{figure*}[htb]
\centering
\includegraphics[width=6.5cm]{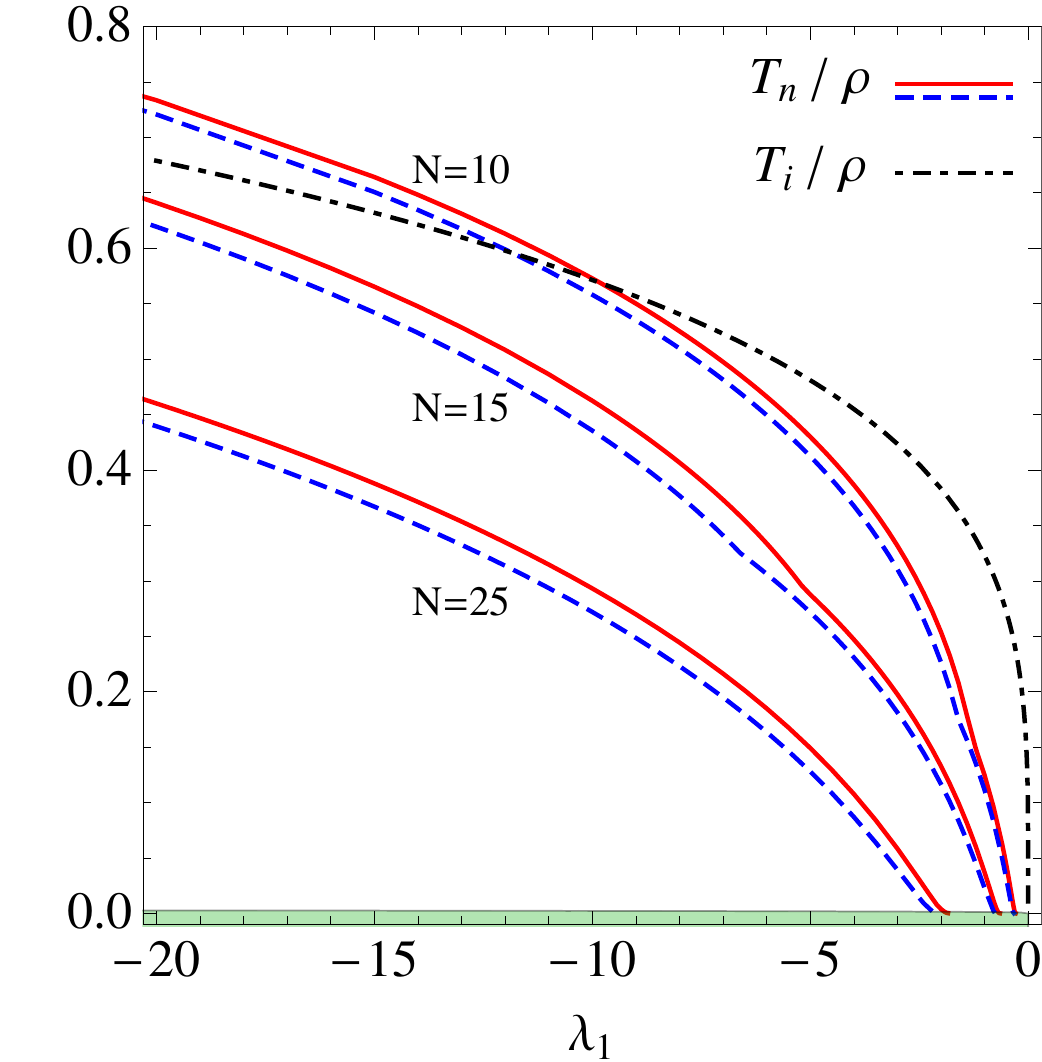}  \hspace{0.9cm} 
\includegraphics[width=6.5cm]{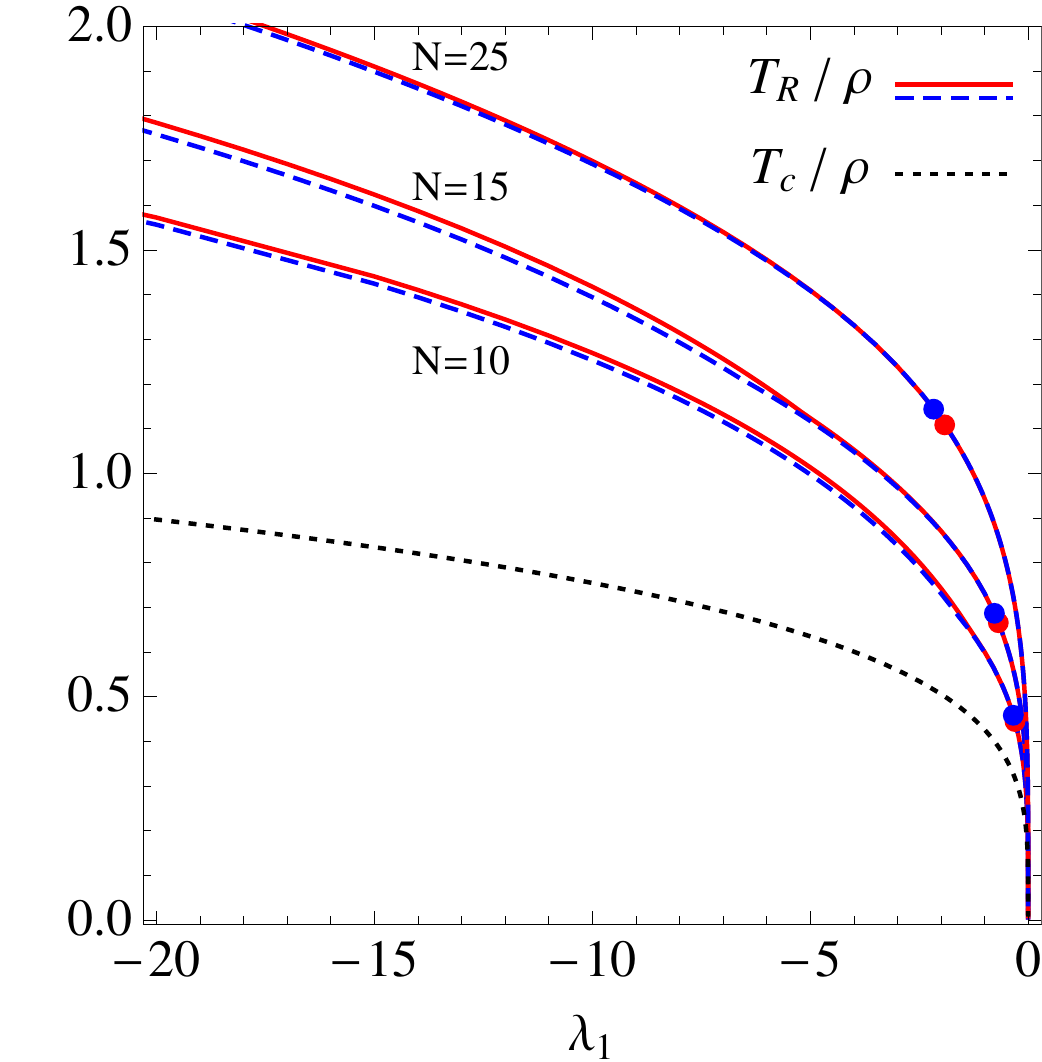} 
\includegraphics[width=6.5cm]{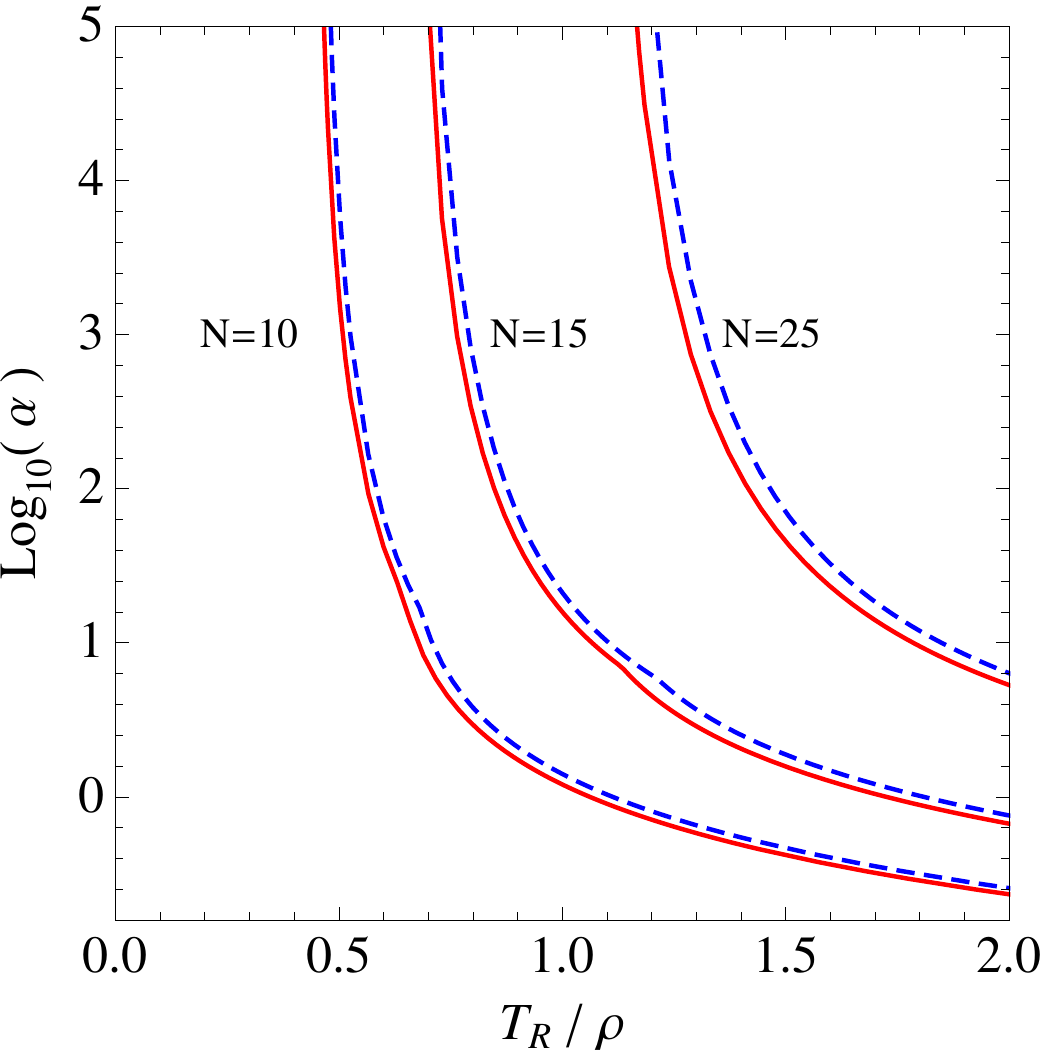}  \hspace{0.9cm} 
\includegraphics[width=6.5cm]{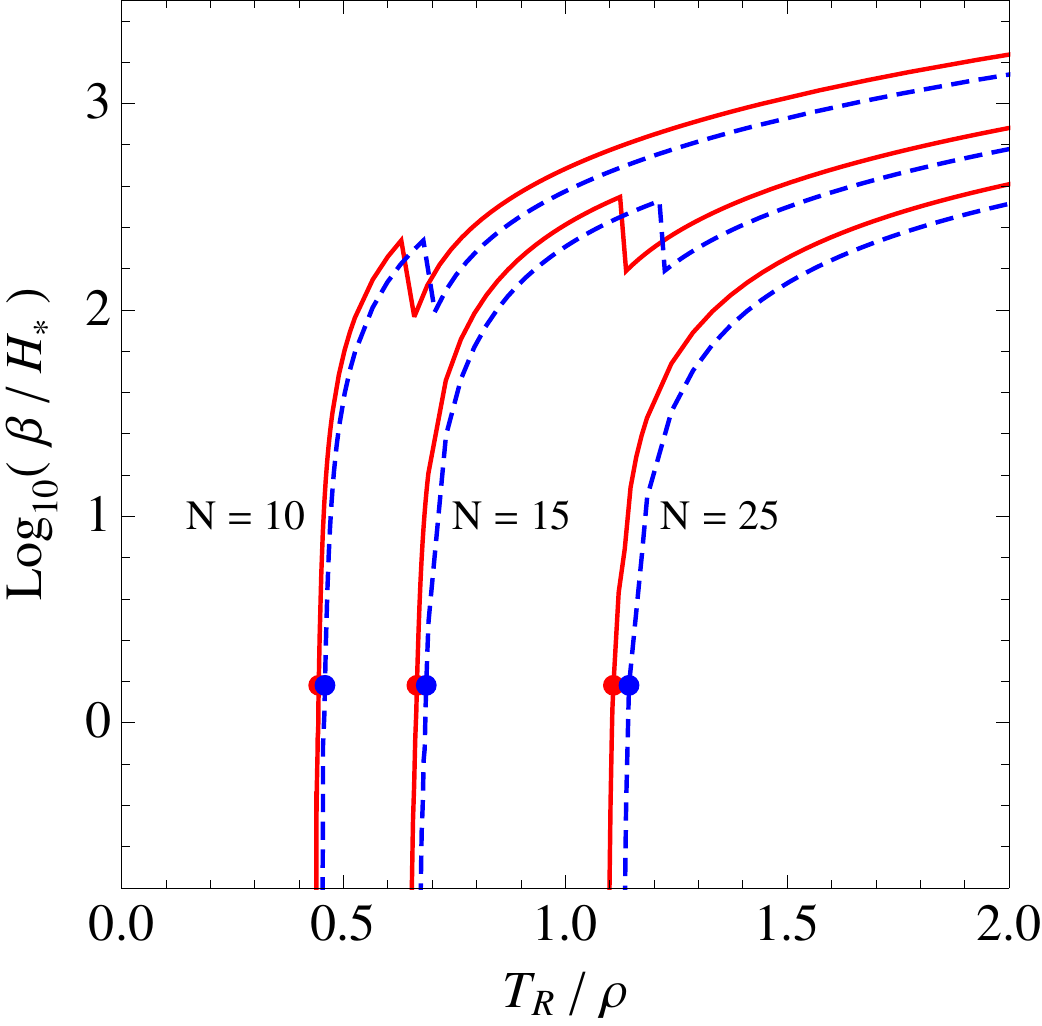} 
\caption{\it Upper panels: $T_n/\rho$ (left panel, solid and dashed
  curves) and $T_i/\rho$ (left panel, dashed-dotted black curve),
  $T_R/\rho$ (right panel, solid and dashed curves) and $T_c/\rho$
  (right panel, dotted black curve) as a function of $\lambda_1$ for
  different values of $N$ and $\rho=1 (100)$\,TeV and the other
  parameters chosen as described in the text. Solid-red (dashed-blue)
  lines correspond to the case $\rho=1$~TeV (100 TeV).  Lower panels:
  $\alpha$ (left panel) and $\beta/H_*$ (right panel) as a function of
  $T_R/\rho$ along the curves displayed in the upper panels with the
  same color and mark. Shadowed (green) region on the bottom on the
  upper left panel is excluded by BBN as $T_n/\rho \gtrsim 3 \cdot
  10^{-4} \sqrt{N}$. The circles correspond to the parameter
  configurations on the border of the BBN bound displayed in
  Fig.~\ref{fig:SNR}.}
\label{fig:Tnlambda1}
\end{figure*}

The Euclidean actions scale as $N^2$ so that they blow up in the limit
$N\to\infty$, where there is no phase transition, so that we focus on
reasonable large values of $N$~\footnote{From the
    AdS/CFT correspondence, the loop expansion on the 5D gravity side
    corresponds to a large number of colors ($N$ large) in the gauge
    theory. In fact from Eq.~(\ref{eq:AdSCFT}) the condition for
    classical gravity to be a good description $M\ell\gtrsim 1$
    translates into the condition $N\gtrsim 4\pi$.}. We find that
$T_n$, in units of $\rho$, has a very mild dependence on $u$, and thus
on $\rho$ itself, whereas it is very sensitive to $N$ and
$\lambda_1$. This is manifest in Fig.~\ref{fig:Tnlambda1} (upper left
panel) which shows $T_n/\rho$ as a function of $\lambda_1$ for
$N=10,15,25$, $\rho=1,100\,$TeV, $\bar v_0 = 1$, $\bar v_1 = 2$ and
$1/\ell = 10^{18}$~GeV (for each value of $\lambda_1$, $u$ is adjusted
to provide $\rho=1,100\,$TeV). The small shift between the dashed and
solid curves precisely comes from varying $\rho$. For all inputs but
$N$ fixed, $T_n/\rho$ decreases with increasing $N$ until reaching a
critical value of $N$ above which the phase transition does not
happen. A similar upper bound on $\lambda_1$ arises when all inputs
but $\lambda_1$ are unchanged.
 
 We stress that $O(4)$ is a good symmetry only for bubbles with critical radius  $R_c < 1/T_n$. This condition is satisfied 
 whenever the $O(4)$ solution dominates, for which we find $R_c T_n \lesssim 0.5$. Moreover, for the SGWB profiles discussed in the next section, the Big Bang Nucleosynthesis (BBN) bound~\cite{Cyburt:2004yc, Caprini:2018mtu} turns out to require in practice $T_n/\rho \gtrsim 3 \cdot 10^{-4} \sqrt{N}$. This lower bound for $T_n/\rho$ is displayed as a shadowed (green) region in Fig.~\ref{fig:Tnlambda1} (upper left panel).
\begin{figure*}[htb]
\centering
\includegraphics[width=6.9cm]{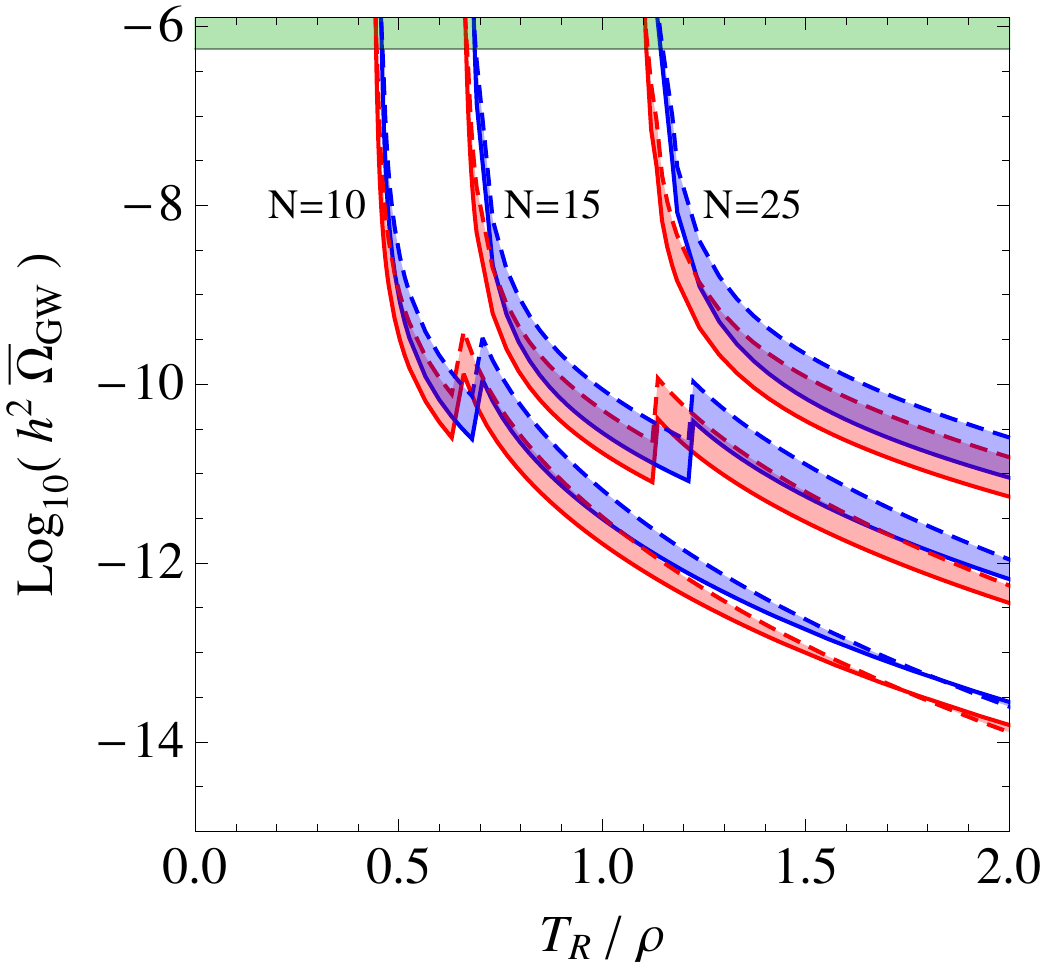}  \hspace{0.9cm} 
\includegraphics[width=6.7cm]{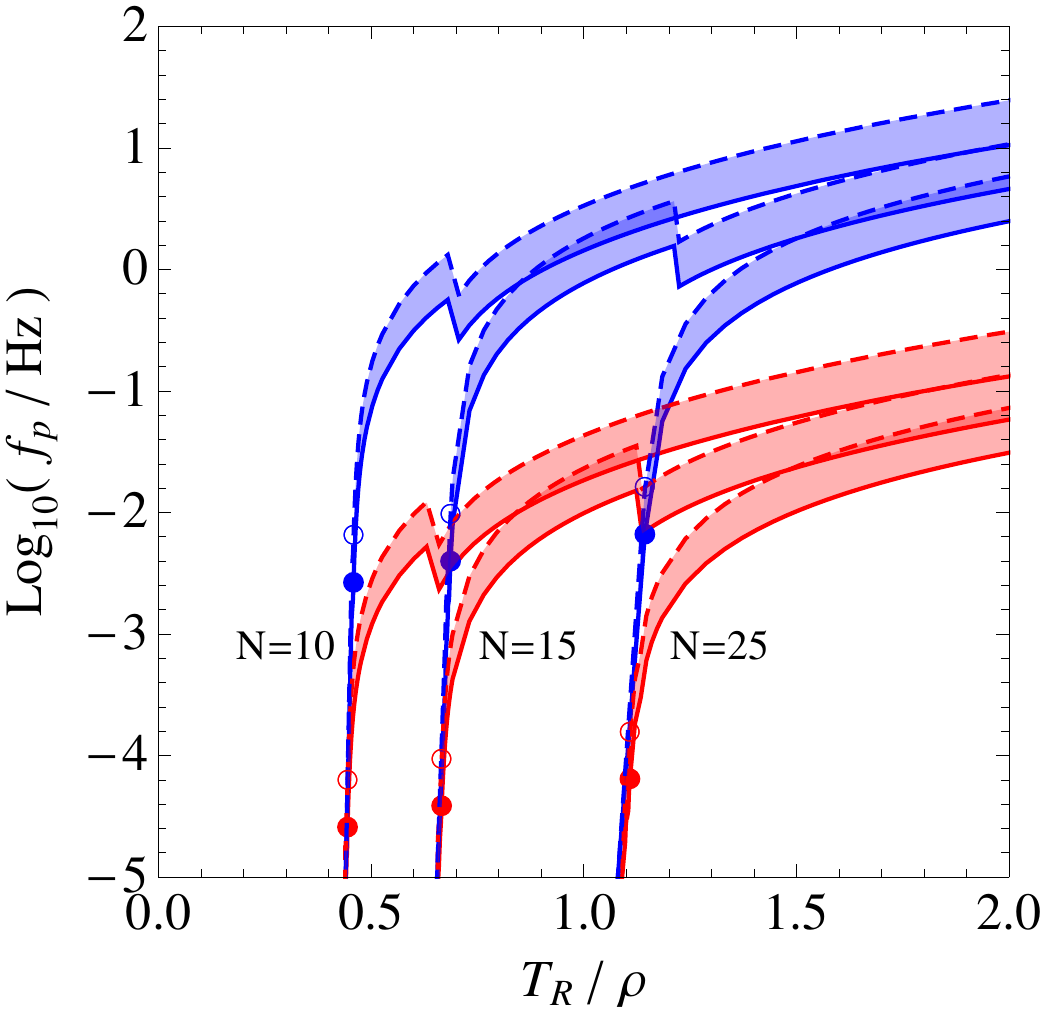} 
\caption{\it Plots of $h^2 \overline\Omega_{\GW}$ (left panel) and $f_p$ (right panel) as functions of $T_R/\rho$ for different values of $N$ and $\rho$. The strips are for $\rho = 1$\,TeV (red) and $\rho = 100$\,TeV (blue). Solid and dashed lines on the edge of the strips correspond to the regime  $\overline\Omega_{\GW}\simeq \overline\Omega_{\GW}^{\rm env}$ and $\overline\Omega_{\GW}^{\rm sw}$, respectively. Shadowed (green) region on the top of the left panel is excluded by BBN. In the right panel the filled circles and the empty circles  correspond to the parameter configurations on the border of the BBN bound for a SGWB profile following Eq.~\eqref{eq:omenv} and Eq.~\eqref{eq:omsw}, respectively. We have considered $v_\omega\simeq 0.99$.}
\label{fig:Omegabar_fp}
\end{figure*} 
The degree of supercooling for nucleation temperatures $T_n\ll T_c$ may trigger a brief period of cosmological inflation.
 We compute the temperature at which inflation starts, $T_i$, by imposing the condition that the energy density in the deconfined phase,
\be
\rho_d(T)=E_0+3\pi^4\ell^3 T^4/\kappa^2+\pi^2g_d^{\rm eff}T^4/30 \label{eq:rhod}
\ee
be dominated by the vacuum energy $E_0$. This gives 
\begin{equation}
T_i\simeq T_c \left[ 3+4 g_{d}^{\rm eff}(T_n)/15N^2  \right]^{-1/4} \,.
\end{equation}
The number of e-folds of inflation produced in the deconfined phase before the transition amounts to
$
N_e=\log(T_i/T_n)
$
provided that $T_i>T_n$. In Fig.~\ref{fig:Tnlambda1} (upper left panel) we plot  $T_i/\rho$ only for $N=25$, while the $N$-dependence of $T_i/\rho$ is tiny. In most of the considered parameter space the supercooling triggers a few e-folds of inflation at most.

After the phase transition, the energy density in the deconfined phase $\rho_d$ is converted into radiation density in the confined phase, and the temperature goes up to the reheat temperature $T_R$.
The requirement $\rho_c(T_R)=\pi^2 g_c^{\rm eff}T^4/30\simeq \rho_d(T_n)$ implies
\be
\frac{4}{15N^2} g_c^{\rm eff}T_R^4=T_c^4+\left(3+\frac{4}{15N^2}g_d^{\rm eff}\right) T_n^4 \,,
\label{eq:TR}
\ee
and the reheating is not huge as Fig.~\ref{fig:Tnlambda1} (upper right panel), which shows $T_R/\rho$ as a function of $\lambda_1$ for various values of $\rho$ and $N$, highlights. We also show in the right panel of Fig.~\ref{fig:Tnlambda1} the plot of $T_c/\rho$ as a function of $\lambda_1$, from its value defined in Eq.~(\ref{eq:tempcritica}), where we see that the condition $T_R>T_c$ is always satisfied.

\section{Gravitational waves}
\label{sec:GW}
A cosmological first order phase transition produces a SGWB whose
power spectrum $\Omega_{\textrm{GW}}(f)$ depends on the dynamics of
the bubbles and their interactions with the
plasma~\cite{Espinosa:2010hh, Caprini:2015zlo, Giese:2020rtr}. When the plasma
effects are negligible, the power spectrum $\Omega_{\textrm{GW}}(f)$ behaves as ~\cite{Huber:2008hg, Weir:2016tov}
\be
\Omega^{\textrm{env}}_{\textrm{GW}}(f)\simeq 
\frac{3.8\, x^{2.8}}{1+2.8\,x^{3.8}}\,\overline\Omega_{\rm GW}^{\,\rm
  env}\,,
  \label{eq:omenv}
  \ee
whereas in the opposite regime it behaves as~\cite{Hindmarsh:2015qta, Hindmarsh:2017gnf, Caprini:2019egz}
\be
\Omega^{\textrm{sw}}_{\textrm{GW}}(f)\simeq x^{3}
\left( \frac{7}{4+3\,x^2} \right)^{7/2} \,\overline\Omega_{\rm GW}^{\,\rm
  sw}\;.
  \label{eq:omsw}
  \ee
(Here $\overline\Omega_{\rm GW}^{\,\rm env, sw}$, $f$, and $f_p$
  stand for the amplitude, frequency, and peak frequency of the power
  spectrum, and $x=f/f_p$.)  The plasma effects are unknown for a
  supercooled radion phase transition (see discussions
  in~\cite{Caprini:2019egz}), hence we use
  $\Omega^{\textrm{env}}_{\textrm{GW}}(f)$ and
  $\Omega^{\textrm{sw}}_{\textrm{GW}}(f)$ to set the theoretical error
  on our SGWB prediction~\footnote{The turbulence
    contribution~\cite{Caprini:2009yp, Kahniashvili:2009mf} and the
    different high-$f$ behavior inferred in~\cite{Cutting:2018tjt}
    fall within this theoretical uncertainty.}.

\begin{figure}[t]
\centering
\includegraphics[width=7.5cm]{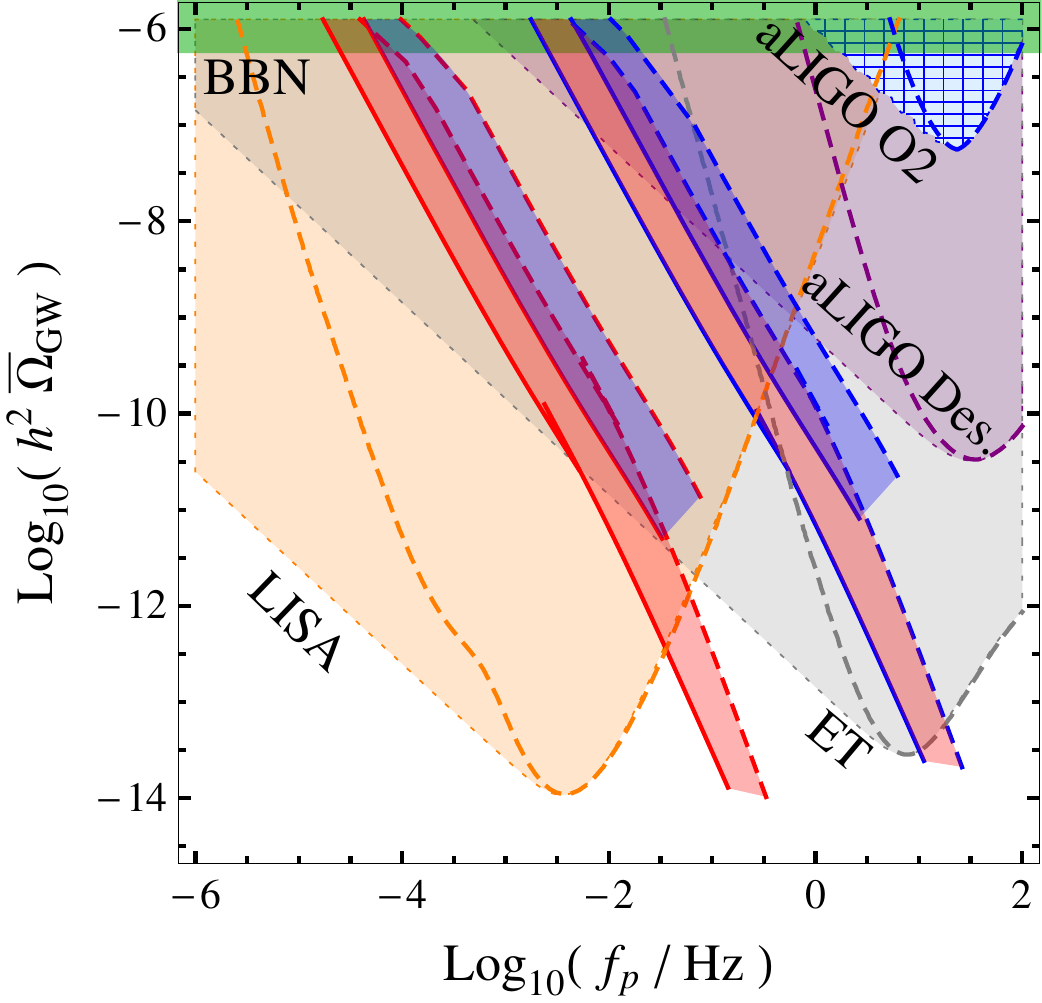}
\caption{\it Parameter reach in the $\{ h^2 \overline\Omega_{\GW},f_p\}$ plane for SGWBs in the regimes $\Omega_{\GW}^{\rm env}$ (regions inside dotted borders) and $\Omega_{\GW}^{\rm sw}$ (regions inside dashed borders). Diagonal strips are for $N=10$ (red) and $25$ (blue) for $\rho = 1$\,TeV (left set) and  $\rho = 100$\,TeV (right set). Solid and dashed lines on the edge of the strips correspond to the regime $\overline\Omega_{\GW}\simeq \overline\Omega_{\GW}^{\rm env}$ and $\overline\Omega_{\GW}^{\rm sw}$, respectively. Regions inside the  areas labeled aLIGO O2 and BBN are in tension with current data.}
\label{fig:SNR}
\end{figure}

For $\overline\Omega_{\rm GW}^{\,\rm env}$, $\overline\Omega_{\rm
  GW}^{\,\rm sw}$, and $f_p$, we use the expressions provided
in Refs.~\cite{Caprini:2015zlo, Caprini:2019egz}.  They depend on:
the normalized gap
between the free energies in the two phases,
\be\alpha = \frac{|F_{d}(T_n) - F_{c}(T_n)|}{\rho_{d}^\ast(T_n)} \,,
\ee
where $\rho_{d}^\ast(T_n) = \rho_d(T_n) - E_0$ is the radiation energy density, cf. Eq.~(\ref{eq:rhod}); the normalized inverse time duration 
\be
\frac{\beta}{H_\ast} = \left. T \frac{dS_E}{dT}\right|_{T=T_n}\, ; \label{eq:betaH}
\ee
 and the wall velocity $v_w$.  The smaller $\beta/H_\ast$ and larger
 $\alpha$, the stronger the phase transition and the SGWB signal. For
 very strong phase transitions one expects $v_w$ much larger than the
 sound speed.  Fig.~\ref{fig:Tnlambda1} (lower panels) shows $\alpha$
 (lower left panel) and $\beta/H_\ast$ (lower right panel) as
 functions of $T_R/\rho$ for the aforementioned input values. The
   stepwise behavior shown in the lower right panel of
   Fig.~\ref{fig:Tnlambda1} (see also Fig.~\ref{fig:Omegabar_fp}), is
   a consequence of the change of regime from $O(4)$ to $O(3)$ bubbles
   when increasing the absolute value of the IR brane parameter
   $\lambda_1$, and correspondingly when increasing the value of
   $T_R/\rho$, cf.~Eqs.~(\ref{eq:SE}) and
   (\ref{eq:betaH})~\footnote{The breaks in the lines of
   $\beta/H_\ast$ vs $T_R/\rho$ for $N=25$ happen at $T_R/\rho \simeq
   2.5$, and so they fall outside the range of
   Fig.~\ref{fig:Tnlambda1} (lower right panel). Notice that similar
   stepwise behaviors in the parameters $h^2 \bar \Omega_{\rm GW}$ and
   $f_p$ appear as well, inherited from the corresponding behavior of
   $\beta/H_\ast$. Consequently, some breaks also appear in the lines
   of Figs.~\ref{fig:Omegabar_fp}-\ref{fig:SNR}, which fall outside
   the range of these figures for $N=25$.}.  From these numerical
 findings we estimate the SGWB signals constituting the theoretical
 predictions on $\Omega_{\textrm{GW}}(f)$ in our setup.

The predicted values of $\Omega_{\rm GW}$ at the peak frequency, and the peak frequency $f_p$ as functions of $T_R$ are shown, respectively, in the left and right panels of Fig.~\ref{fig:Omegabar_fp} for $\rho=1$ TeV (red strips) and $\rho=100$ TeV (blue strips), and for different values of $N$. The borders of the strips marked in solid are evaluated by means of Eq.~\eqref{eq:omenv} for $v_w=0.99$ and the values of other phase transition parameters  displayed in Fig.~\ref{fig:Tnlambda1}. The borders marked in dashed are evaluated in the same way but by means of Eq.~\eqref{eq:omsw}.  The strips can thus be interpreted as the model predictions and their uncertainties. In the
parameter space $\{ \overline{\Omega}_{\rm GW} , f_p\}$, such strips translate to those reported in Fig.~\ref{fig:SNR}.  
They are cut in their lower part when $T_R \lesssim 2 \rho$ which also yields $|\delta|/ \bar{r}_1^m \lesssim 0.1$.  This prevents the (large) detuning from jeopardizing our perturbative expansions and suppresses the heavy Kaluza-Klein resonances to be in thermal equilibrium in the relativistic plasma. We remark that the strips are displayed for $v_w=0.99$ but no significant change would be visible for e.g.~$v_w\simeq 0.7\,$. 

Fig.~\ref{fig:SNR} includes the sensitivity prospects to the
parameter space $\{ \overline{\Omega}_{\rm GW} , f_p\}$ in the
$\Omega^{\textrm{env}}_{\textrm{GW}}$ and
$\Omega^{\textrm{sw}}_{\textrm{GW}}$ approximations (shadowed areas within dotted and dashed borders, respectively). It forecasts the situation
expected towards the end of the next decade when LISA, ET and aLIGO
will have run for several years. For concreteness, the sensitivity
regions assume 3, 7 and 8 years of usable data for LISA, ET and aLIGO Design,
respectively~\cite{Audley:2017drz, Sathyaprakash:2012jk, Aasi:2013wya}. The
exclusion bounds from BBN~\cite{Cyburt:2004yc, Caprini:2018mtu} and present searches in aLIGO O2~\cite{LIGOScientific:2019vic}  are also
recast.  The former varies very little in one or the other SGWB approximation, namely $h^2 \overline{\Omega}_{\rm GW}^{\rm env} < 5.6 \cdot 10^{-7} $ and $h^2 \overline{\Omega}_{\rm GW}^{\rm sw} < 7.7 \cdot 10^{-7} $. 
The aLIGO O2 bound comes from null searches for signals with signal-to-noise ratio
SNR$\,\ge 2$~\footnote{Technically, the aLIGO analysis only excludes scenarios with a
power law SGWB. The analysis is however not blind to slightly
more powerful SGWBs with a less-trivial frequency shape, although the confidence level of the exclusion remains to be quantified.}. Thus, we also forecast  the future sensitivity regions by
adopting the same criterion, SNR$\,\ge 2$, and implementing the noise curves provided in the official documents~\cite{Audley:2017drz, Sathyaprakash:2012jk, Aasi:2013wya} (see Refs.~\cite{Thrane:2013oya, Caprini:2015zlo, Karnesis:2019mph, Caprini:2019pxz, ptplot, Schmitz:2020syl} for shortcuts and other approaches).

The forecast shows that by the late 2040s the planned interferometer
network will be sensitive to a huge parameter region, probing $\rho$
up to the $10^{9}$-TeV scale.  Thanks to the complementary of the
network, the theoretical uncertainty on the plasma effects during the
transition only marginally affects the parameter reach of the whole
network.  The uncertainty is instead relevant in the next years when
only aLIGO operates, e.g.~aLIGO Design reaches scenarios with
$\rho\sim 100\,$TeV only in the
$\Omega^{\textrm{env}}_{\textrm{GW}}(f)$ regime.

The broad parameter reach emerging from the forecast is promising not only in terms of detection but also of reconstruction. In fact, most of the benchmark scenarios fall well inside the sensitivity regions shown in Fig.~\ref{fig:SNR}. These scenarios hence exhibit SGWBs with a large SNR~\footnote{Parameter points on the borders of the sensitivity regions have $\textrm{SNR} = 2$ by construction, and the SNR scales with $\overline{\Omega}_{\textrm{GW}}$ at a given $f_p$. }, and a sizable SNR typically implies small uncertainties on the signal reconstruction~\cite{Caprini:2019pxz}. Of course, this SNR argument relies on the size of the SGWB signal relatively to the instrumental noise, but in general fails if the phase transition signal co-exists with other powerful SGWBs sources.  Among the feasible SGWBs, those of astrophysical origin \cite{Nelemans:2001hp, Ruiter:2007xx, Sesana:2016ljz, LIGOScientific:2019vic, Bonetti:2020jku} are dominant only at the margins of the forecast sensitivity regions. In addition, in the late 2040s it will be possible to dig out signals much weaker than these astrophysical backgrounds if their templates are accurate enough~\cite{Pieronietal}. Instead,  SGWBs of cosmological origin can potentially be  problematic. For instance, in some extreme setups, SGWBs sourced by inflationary phenomena or cosmic strings can be as powerful as the strongest signals predicted in our model, however the plausibility of such extreme setups is doubtful. It seems then likely that the reconstruction, and the subsequent parameter estimation, of (most of) the signals predicted in the considered warped model will be accurate.

We finally remark that at the qualitative level our findings should apply to any warped setups with radion stabilization mechanism. Such setups are indeed expected to have a SGWB phenomenology similar to that here studied. In this sense, our findings show that aLIGO O2 data already corners vanilla warped scenarios with $\rho \sim 10^5\,{\rm TeV} $ and extremely strong phase transitions.

\section{Conclusions}
\label{sec:conclusions}
We have analyzed warped models with the radion stabilized by 
a polynomial potential, in the regimes of small and sizable backreaction. As the backreaction is an important ingredient to generate an effective potential with a stable minimum, we have conveniently used the superpotential method to analytically tackle with it. However, to zeroth order in the superpotential $s$-expansion, the superpotential method is not a good tool to generate an effective potential, an observation already done in Ref.~\cite{Megias:2018sxv}, as it simply yields a runaway behavior. We have then worked out to first order in the $s$-expansion, using techniques previously introduced in Ref.~\cite{Konstandin:2010cd}, a self consistent method if working in a region of the parameter space where the $s$-expansion converges fast, as we have proved throughout this paper. In the region of small backreaction we have found good agreement with previous results in the literature, as the original Goldberger-Wise potential from Ref.~\cite{Goldberger:1999uk}. Moreover, as the presence of new physics has been elusive up to now, we have considered the possibility of heavy Kaluza-Klein resonance masses, thus leaving open the door that nature has chosen to provide our particle physics model with a severe little hierarchy problem and the corresponding level of fine-tuning.

Using the radion zero-temperature effective potential shaped by the
backreaction, and standard techniques of 5D warped theory at finite
temperature, we have studied the radion (confinement/deconfinement)
first order phase transition and the stochastic gravitational wave
background that such a phase transition generates. We have then
compared the obtained gravitational wave signatures with the
corresponding detection capabilities of present (aLIGO) and future
(ET, LISA) gravitational wave interferometers.  We have found that in
the next decade the gravitational wave detectors will broadly probe
warped models.

We expect our results to be rather generic. Indeed the radion phase transition of the considered model is similar to the one of many other warped setups of the literature. This implies that in all these models the region with KK resonances at  $m_{\rm KK} \sim \mathcal{O}(10^{5})\,$ -- $\mathcal{O}(10^{6})\,$TeV is being cornered by current aLIGO O2 data. Moreover, the forthcoming interferometers will broadly test these models by being capable to probe resonances of mass $m_{\rm KK} \lesssim 10^5$~TeV~(LISA), $10^2~\textrm{TeV} \lesssim m_{\rm KK} \lesssim 10^8$~TeV~(aLIGO Design) and $m_{\rm KK} \lesssim 10^9$ TeV (ET). In this sense, the future gravitational wave detectors have the great potential to shed light on the little hierarchy problem and the amount of tuning that is acceptable in nature.

\begin{acknowledgments}
\noindent
The authors thank the ICTP South American Institute for Fundamental
Research (SAIFR), Sao Paulo, Brazil, and its Program on Particle
Physics, September 30-November 30, 2019, where part of this work was
done, for hospitality.  The work of EM is supported by the Spanish
MINEICO under Grant FIS2017-85053-C2-1-P, by the FEDER/Junta de
Andaluc\'{\i}a-Consejer\'{\i}a de Econom\'{\i}a y Conocimiento
2014-2020 Operational Programme under Grant A-FQM-178-UGR18, by Junta
de Andaluc\'{\i}a under Grant FQM-225, and by Consejer\'{\i}a de
Conocimiento, Investigaci\'on y Universidad of the Junta de
Andaluc\'{\i}a and European Regional Development Fund (ERDF) under
Grant SOMM17/6105/UGR. The research of EM is also supported by the
Ram\'on y Cajal Program of the Spanish MINEICO under Grant
RYC-2016-20678. The work of MQ is partly supported by Spanish MINEICO
under Grant FPA2017-88915-P, by the Catalan Government under Grant
2017SGR1069, and by Severo Ochoa Excellence Program of MINEICO under
Grant SEV-2016-0588.
\end{acknowledgments}

\bibliography{refs}

\end{document}